\documentclass{article}

\usepackage{amssymb,amsmath,amsthm,color,graphicx,times,graphicx}
\usepackage{hyperref}
\usepackage{enumerate}
\usepackage{subfigure}
\usepackage[affil-it]{authblk}

\begin{document}

\title{Security against jamming  and noise exclusion in imaging}

\author{ Wojciech Roga and John Jeffers\footnote{Corresponding author: wojciech.roga@strath.ac.uk}}

\affil{ SUPA, Department of Physics, University of Strathclyde, John Anderson Building, 107 Rottenrow, Glasgow G4 0NG, UK}



\maketitle

\begin{abstract}
We describe a protocol by which an imaging system could be protected against jamming by a malevolent party. Our protocol not only allows recognition of the jamming, but also allows for the recovery of the true image from the jammed one. We apply the method to jamming of quantum ghost imaging, for which the jamming detection probability is increased when the imaging light is entangled. The method can also be used to provide image recovery in general noisy environments. 
\end{abstract}

\section{Introduction}

Image security is a challenge that arises when spatial information about an object is transferred to and imaged by a device at a remote location. It is more important when receiving particular images implies related actions. Security cannot normally be guaranteed along the entire path between the source of light and the detectors. Imagine a situation in which an intruder places a false object in front of the real one, or alternatively changes the path of a beam to direct it via a false object. There is no protection from such tampering. 
Thus when we simply point a camera we can not be sure whether or not the light received bears any relation to any real object. Someone may simply send us the image that they want us to see. 
Of course we can provide more confidence by controlling of the source of radiation as well as the detector, as is done in radar or lidar systems, but the fundamental point is that imaging is insecure.

Quantum protocols have been applied to 
imaging and communication, providing several advantages \cite{Pittman1995,Kolobov1999,Lugiato2002,Giovannetti2004,Santivanez2011,Jiang2013}. Among them are detection of imaging jamming and appropriate protection~\cite{Malik2012,Humble2008}.
In communication situations we can use encryption to safeguard information in transit. This typically relies on secure key-sharing protocols, an example of which is quantum key distribution ~\cite{Bennett1984,Ekert,Lutkenhaus2000,Sabottke2012}. When this procedure succeeds two parties share a secrete key and only limited private information can leak to an eavesdropper. In imaging, however, the privacy condition is not normally crucial. We simply care that the image received exactly corresponds to the object. This is an example of public communication in which we stress the correctness of the message or image instead of its privacy. We face not an eavesdropper but an active intruder, who may already know the object and, instead, wants to jam the imaging protocol, changing information and consequently perhaps a strategic decision.

As already stated, there is no secure protection from an intruder who uses a false object or a false beam path. It may, however, be difficult for the intruder to perform such jamming both efficiently and undetectably. Thus we investigate a weaker jamming procedure in which some of the light from a trusted source of illumination is intercepted by the intruder and replaced by light from their independent source \cite{Humble2009,Bennett1992,Gisin2002}. As the party who prepares states of light has naturally more information about the states than it is possible to extract from a measurement, the legitimate imager has an informational advantage over the intruder. This advantage can be a source of imaging security as we show in this paper. 

The paper is organised as follows. In Sec. \ref{sec2}, we analyse the general image security scenario and provide a universal description of intruder detection and correct image recovery. Our protocol applies to both deliberate jamming and to image noise. In Sec. \ref{sec3}, we define a quantitative criterion for image comparison that enables us to estimate the probability that we detect jamming and a false alarm probability. In Sec. \ref{sec5}, we propose an arrangement of ghost imaging~\cite{Pittman1995,Erkmenn2010,Bennink2002,Chan2009,Aspden2015} protected against such jamming by using both classical and quantum-correlated light. Here, we show an advantage of using the quantum-correlated light and demonstrate the recovery protocol.
Finally, concluding remarks and possible extensions of these work are discussed in Sec. \ref{sec6}.  



\section{Universal intrusion detection and image recovery protocol} \label{sec2}
Let us start with a general description of imaging protected by a security system that allows us to both detect the presence of an intruder and to counter the effect of jamming. We note at the outset that there is nothing specifically quantum about our procedure. It can be performed with classical light. There can sometimes be advantages, however, to using certain types of quantum state.

We assume that states of light $\rho({\bf x},\kappa)$ used to image an object are characterized by spatial coordinates ${\bf x}$ describing the coordinates of light during the procedure together with another degree of freedom $\kappa$ that is used to detect intrusion and recover the image. In this paper $\kappa$ denotes a polarisation state, however all the results could apply to other degrees of freedom. 
Suppose that states produced by a trusted source interact with an object and therefore carry information about it. These states are denoted $\rho_j({\bf x},\kappa)$, where $j$ indicates that the legitimate imager can choose from a set containing more than one state of light for imaging. Indeed, this diversification is crucial to the protocol.

An intruder intercepts a fraction $r$ of the light and resends their own photons in state $\rho^E({\bf x},\kappa')$. Superscript $E$ refers to an intruder or jammer or simply a noisy environment that introduces errors. We assume that the state selected by the intruder does not depend on the choice of $j$ made randomly by the imagers and is, at least on average, constant in time. This is a reasonable assumption because in imaging the dynamics of the process is not typically relevant. We will discuss this condition further in the context of particular implementations. In the final part of the paper we comment on the situation in which it is partially relaxed, i.e. the state of the intruder changes in time in such a way that a partial correlation between $\rho^E({\bf x},\kappa')$ and $j$ is established. 

The intrusion detection and image recovery arrangement contains a set of polarisation analysers. The analysing system induces the detectors to make a measurement of $\kappa$ by filtering,  characterised by a set of parameters $\{\theta_i\}$, which in our case denotes the set of angles of different polarisers. We consider a general situation of multi-photon coincidence imaging with $n$ photons involved in which $i=1,...,n$. A spatial distribution of intensities in an image is proportional to probabilities corresponding to an $n$-photon polarisation state $\rho=\rho({\bf x},\kappa)$ which depend on the analyser angles $\theta_i$ and are given by
\begin{equation}
 P\left(\{\theta_i\},\rho\right)=\Big(\langle a(\theta_1)|\otimes...\otimes \langle a(\theta_n)|\Big)\rho\Big(|a(\theta_1)\rangle\otimes...\otimes |a(\theta_n)\rangle\Big).
 \label{pleft}
\end{equation}
This is a probability if $\rho$ is a state with exactly n-photons, otherwise it is an expectation value. Here, for a single analyser and fixed basis of horizontal-vertical polarisations, $|a(\theta_i)\rangle=\cos{\theta_i}|\!\leftrightarrow\rangle+\sin{\theta_i}|\!\updownarrow\rangle$. 
The state of the entire system, before any action of the analysers
can be written as $(1-r)\rho_j+r\rho^E$. The intensity distributions of images are proportional to
\begin{equation}
(1-r)P\left(\{\theta_i\},\rho_j\right)+r P\left(\{\theta_i\},\rho^E\right).
\end{equation}
Here several images indexed by $j$ and $i$ are obtained. If the intensity dependence of the received images corresponds to the distribution $P\left(\{\theta_i\},\rho_j\right)$, which is known only to the imager, then we can infer that there is no intrusion. If not, intrusion is detected and, if it is only partial ($r<1$), the correct image can be recovered, as we now show.  

As an example assume that imager uses two different states of polarisation, so $j=1, 2$. The contribution of the intruder $\rho^E$ does not depend on $j$ so the detected images $(1-r)P\left(\{\theta_i\},\rho_1\right)+r P\left(\{\theta_i\},\rho^E\right)$ and $(1-r)P\left(\{\theta_i\},\rho_2\right)+r P\left(\{\theta_i\},\rho^E\right)$ have the same contribution to the image induced by the intruder. If the contribution from $P\left(\{\theta_i\},\rho_1\right)$ is different to the contribution from $P\left(\{\theta_i\},\rho_2\right)$ the absolute value of the difference between the two images for any $i$ eliminates any incorrect part of the image while merely attenuating any correct part because the difference of the images is taken. This procedure allows us to distill a correct, although attenuated, image of the investigated object.

\section{Quantitative comparison of images}\label{sec3} 
\subsection{Probability of jamming detection}
An essential part of the detection protocol is based on the ability of legitimate imagers to distinguish between images formed by two kinds of state characterised by different polarisation degrees of freedom. In order to compare images quantitatively we define a {\it state dependent visibility} $V$ that for given pair of polarisation states $\rho_1$ and $\rho_2$ and for a given set of analyser orientation angles $\{\theta_i\}$ is given by 
\begin{equation}
V(\rho_1,\rho_2,\{\theta_i\})=\frac{|P_1-P_2|}{P_1+P_2},
\end{equation}
where $P_1$ and $P_2$ are photon detection probabilities for states $\rho_1$ and $\rho_2$ respectively as in (\ref{pleft}). Although, for brevity, we omit the arguments of $P_j$ in our notation, we assume that $P_j=P_j\left(\{\theta_i\},\rho_j\right)$ i.e. $P_j$ is defined for a specific state and for a chosen set of analyser angles. Notice that for $\rho_1=\rho_2$ the state dependent visibility vanishes, showing that the images are identical. If, instead, for two different states $V\neq 0$ it means that intensity distributions of two images are different.

As the legitimate imagers know both the states used and the analyser orientations in ideal conditions without intrusion they expect to observe images of a particular state dependent visibility. Any noise or intrusion will reduce the observed visibility. 

The measured and expected visibilities provide data for hypotheses testing based on the likelihood ratio test~\cite{VanTrees2001,Humble2009}. Our null hypothesis $H_0$ assumes no intruder while the alternative hypothesis $H_1$ assumes the presence of intercept-resend jamming. We must also account for real devices having some level of noise. Let us assume that the measured visibilities are associated with the Gaussian noise of variance $\sigma$ and test the hypotheses based on the likelihood ratio test with prior probabilities $0.5$ (and the test threshold $\lambda=1$). Visibilities related to the two hypotheses are the following: for 
$H_0$ we have $s_0=V_0+N$ and for $H_1$, $s_1=V_1+N$ where $V_0$ and $V_1$ are the average values of the visibilities and $N$ is the Gaussian noise with variance $\sigma$. In the log-likelihood ratio test we decide that $H_0$ occurs if
\begin{equation}
 \sum_i\tilde{s_i}<\frac{{\rm ln}\lambda}{d}+\frac{d}{2},
 \label{test}
\end{equation}
where $\tilde{s_i}=s_i/(\sigma \sqrt{M})$, and $M$ is a number of trials which we will assume to be 1. Here $d=\sqrt{M}(V_1-V_0)/\sigma$. A corresponding inequality to Ineq.~(\ref{test}) convinces us to accept $H_1$. The probability of correct detection of an intruder (i.e. when $H_1$ is correctly accepted) and the false alarm probabilities are
\begin{eqnarray}
 P_d&=&{\rm erfc}\left[\frac{{\rm ln}\lambda}{d}-\frac{d}{2}\right],  \label{pede}\\
P_{\rm err}&=&{\rm erfc}\left[\frac{{\rm ln}\lambda}{d}+\frac{d}{2}\right],
 \label{perr}
\end{eqnarray}
and ${\rm erfc}[x]$ denotes the complimentary error function~\cite{VanTrees2001}. 

\subsection{Application of the detection strategy, level of jamming} \label{sec4}
In order to demonstrate the optimal decision strategy made by the legitimate imagers let us consider the attack in which the intruder intercepts a fraction $r$ of the original photons and replaces them with photons in a state $\rho^E$ carrying the false image.

As the intruder detection probability (\ref{pede}) is a monotonic function of the difference between expected and observed state dependent visibilities, the intruder will choose states carrying false information such that this difference is as small as possible. The legitimate imagers have freedom in the choice of analyser angles and will want to maximise this difference. 
In principle, the intruder could optimize their state for each value of $\{\theta_i\}$. Although it is unlikely, the legitimate imagers can consider this situation in order to check the robustness of their state choice against jamming. Thus they are able to decide the states for which the detectability of an intruder is largest. Hence in the worst case scenario the detection probability (\ref{pede}) is a monotonic function of 
\begin{equation}
d=\frac{1}{\sigma}\max_{\{\theta_i\}}\Big[\min_{\rho^E}\Big(V(\rho_1,\rho_2,\{\theta_i\})-V(\rho_1',\rho_2',\{\theta_i\})\Big)\Big],
\label{dworst}
\end{equation}
where $\rho_1$ and $\rho_2$ are the states chosen by the legitimate imagers and 
\begin{equation}
\rho_j'=(1-r)\rho_j+r\rho^E,\qquad {\rm where}\qquad j=1,2
\label{rhoi}
\end{equation}
are the states carrying false information. The parameter $d$ is estimated for a given level of intrusion. In cryptography a similar parameter is given by a secrecy function \cite{Han2014}. 
For image comparison we introduce the {\it level of jamming} in terms of the state dependent visibility for jammed and non-jammed images
\begin{equation}
V_L=\max_j\ V(\rho_j,\rho_j',\{\theta_i\}),
\label{loj}
\end{equation}
where $\rho_j$ means the state from legitimate source and  $\rho_j'=(1-r)\rho_j+r\rho^E$ its jammed counterpart. The level of jamming $V_L$ quantitatively describes  how at most the actual image can differ from the correct one in visibility. In principle, as a measure of the intrusion level we could consider the intercepting rate $r$. However, this is not the optimal choice, as even for large $r$ the correct image may be attenuated but not changed as is shown in the following example.
\vspace{3pt}

\noindent{\it Example. One photon states.} Imagine that the states chosen by legitimate parties are $\rho_1=|\leftrightarrow\rangle\langle\leftrightarrow|$ and $\rho_2=|\updownarrow\rangle\langle\updownarrow|$. We assume that $\rho^E$ can be an arbitrary one photon polarisation pure state $\rho^E=|\phi\rangle\langle\phi|$, where $|\phi\rangle\!=\!\cos\alpha|\leftrightarrow\rangle+e^{i\beta}\sin\alpha |\updownarrow\rangle$, for arbitrary real $\alpha$ and $\beta$. It is easy to show that in this case
\begin{equation}
V(\rho_1',\rho_2',\theta)=\frac{(1-r)V(\rho_1,\rho_2,\theta)}{1-r+r\langle a(\theta)|\rho^E|a(\theta)\rangle},
\end{equation}
where $\rho_i'$ are given by 
\begin{equation}
\rho_j'=(1-r)\rho_j+r\rho^E,\qquad {\rm where}\qquad j=1,2.
\end{equation}
Thus for each $\theta$ the intruder can find a state $\rho^E$ orthogonal to $|a(\theta)\rangle$. Hence, for any intercepting rate $r$ the parameter $d=0$ and the probability of detection is as small as possible, equal to the probability of an unbiased guess $0.5$. However, notice that although the interference of the intruder is significant, the image is not changed apart from being attenuated. 
Because we stress the correctness of the image rather than its intensity, which we do not fully control, this invasion is not treated as jamming. Therefore, we do not treat $r$ as a proper characterisation of the degree of jamming.

\section{Ghost imaging jamming}\label{sec5} 

In ghost imaging correlations between photons in two separate light beams, which we call the object beam and the image beam, allow an image of an object to be obtained by detecting light that has never interacted with it directly. The object beam interacts with an object with a particular spatial profile. This beam is detected with a bucket detector that provides no spatial information. The image beam does not interact with the object but falls on a spatially resolving detector. An image appears at this detector due to coincidence correlations between two detectors. The technique has applications for example, for difficult to access objects where a single pixel detector might be easier to place, or if it is easier to detect spatial images at one wavelength rather than another~\cite{Chan2009,Aspden2015}. The correlations also provide timing information. 

\subsection{Probability of jamming detection in ghost imaging}

In what follows we describe the detection and recovery protocol applied to ghost imaging. This two-photon coincidence imaging allows us additionally to show that imaging with entangled states can increase the probability of intrusion detection compared to classically correlated states.

\begin{figure}[!h]
\begin{center}
\includegraphics[scale=0.4]{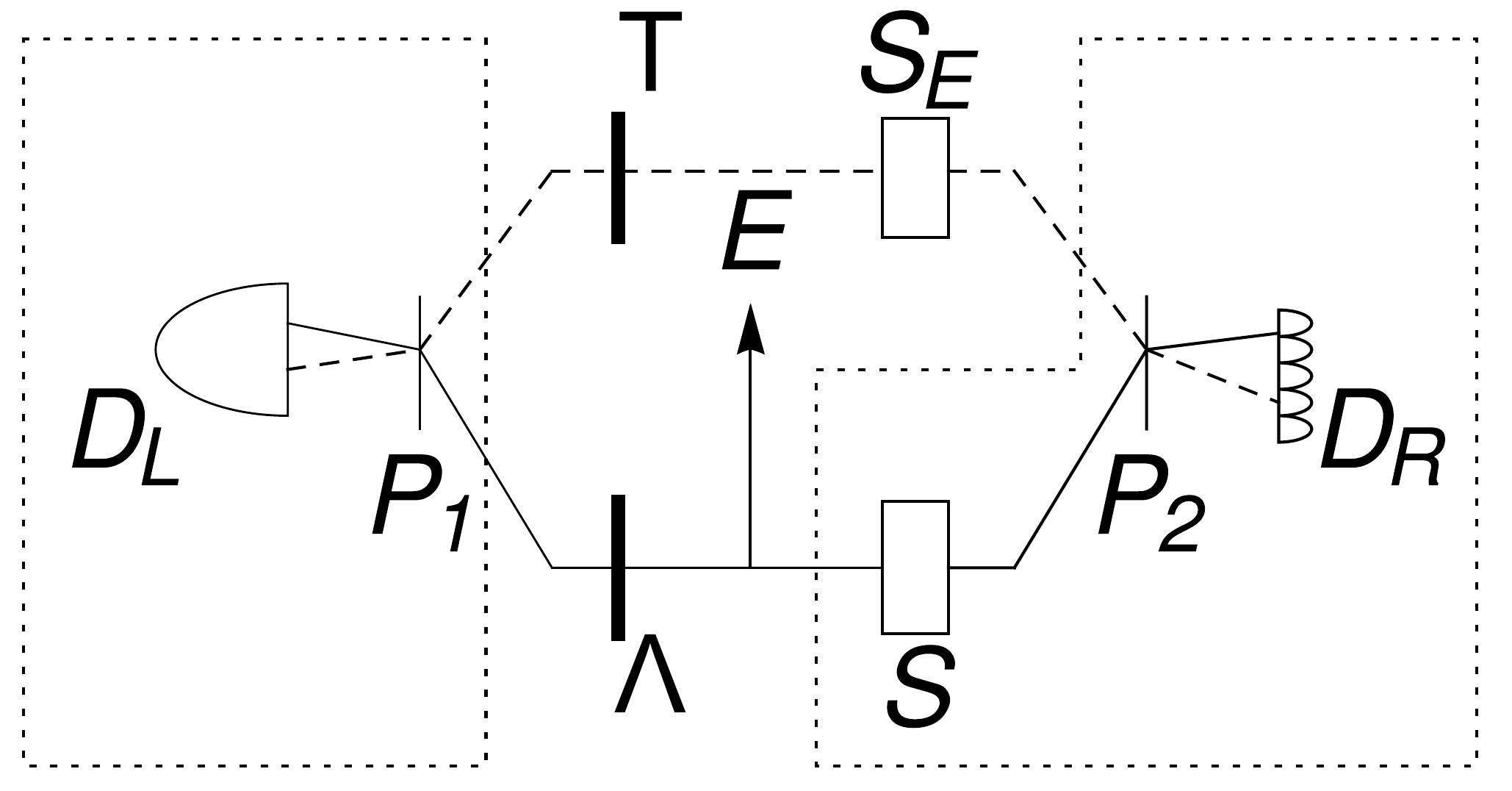}
\caption{Ghost imaging with security. Intruder $E$ intercepts part of a signal emitted by source $S$ sent to an object $\Lambda$. $E$ re-sends correlated pairs of photons produced by source $S_E$ that carry information about a false object $T$ to both right and left detectors. Regions bordered by dotted lines denote zones controlled by imagers.}
\label{img1b}
\end{center}
\end{figure}

A ghost imaging setup contains the following fundamental elements: a source of correlated photons, $S$, the investigated object, $\Lambda$, a bucket detector $D_L$ and a spatially-resolving detector $D_R$. These elements are shown in Fig.~\ref{img1b} which is arranged in the so-called Klyshko picture~\cite{Strekalov1995,Tan2003}. The central part of the setup is doubled, showing the symmetric role of imagers (lower) and an intruder (upper). 
Polarisation analysers $P_1$ and $P_2$ can be rotated independently to distinguish two angles $\theta_1$ and $\theta_2$ of polarisation filtering. Intruder $E$ intercepts a fraction of the photons sent by trusted source $S$ and resends pairs of photons, correlated in polarisation, which carry false image information. We assume that different photons from each pair travel through different arms and $E$ intercepts photons from only one arm. This is enough to prevent coincidence counting which can contribute to a correct image. 

The states used by the legitimate imagers are changed at random. We assume that the intruder does not know which of them was chosen, as the states have maximally mixed reduced states and cannot be distinguished by any local measurement. Therefore, the states carrying the false part of the image the intruder resends are independent of the choice by the legitimate imagers. For given angles of the analysers, legitimate imagers observe two images that have the same contribution from the false image. The difference between these images allows them to recover the correct image.

As we have already mentioned, the legitimate imagers use states with maximally mixed reduced states. All two-photon polarisation states with maximal mixed reduced states can be transformed by local changes of bases to the following class known as Bell diagonal states \cite{Dakic2010,Horodecki1996}
\begin{equation}
\rho_{BD}=\frac{\bf 1}{4}+\mu_x\sigma_x\otimes\sigma_x+\mu_y\sigma_y\otimes\sigma_y+\mu_z\sigma_z\otimes\sigma_z,
\label{BellDiag}
\end{equation}
where $\sigma_x,\sigma_y$ and $\sigma_z$ are three Pauli matrices, ${\bf 1}$ is the identity matrix, $\mu_x,\mu_y$ and $\mu_z$ are real parameters for which (\ref{BellDiag}) is positive semidefined. 
Examples of these states, that will be used are the following: 
\begin{eqnarray}
&&\!\!\!\!\!\!\!\!\!|\psi_1\rangle\!\!=\!\!\frac{1}{\sqrt{2}}(|\leftrightarrow\leftrightarrow\rangle+|\updownarrow\updownarrow\rangle)\ \ \ \quad{\rm for}\ \mu_x\!=\!\mu_y\!=\!\mu_z\!=\!1/4 ,
\label{mes1}\\
&&\!\!\!\!\!\!\!\!\!|\psi_2\rangle\!\!=\!\!\frac{1}{\sqrt{2}}(|\!\leftrightarrow\leftrightarrow\rangle\!-\!|\updownarrow\updownarrow\rangle\!)\ \ \quad{\rm for}\ \mu_x\!=\!\mu_y\!=\!-\frac{1}{4}, \mu_z\!=\!\frac{1}{4} , 
\label{mes2}\\
&&\!\!\!\!\!\!\!\!\!\omega_1\!\!=\!\!\frac{1}{2}(|\!\!\updownarrow\updownarrow\rangle\langle\updownarrow\updownarrow\!|\!+\!|\!\!\leftrightarrow\leftrightarrow\rangle\langle\leftrightarrow\leftrightarrow\!\!|)\ {\rm for}\ \mu_x\!=\!\mu_y\!=\!0, \mu_z\!=\!\frac{1}{4} ,\ \ \ \ \ 
\label{ccs1}\\
&&\!\!\!\!\!\!\!\!\!\omega_2\!\!=\!\!\frac{1}{2}(|\!\!\searrow\hspace{-10pt}\nwarrow\!\searrow\hspace{-10pt}\nwarrow\!\rangle\langle\!\searrow\hspace{-10pt}\nwarrow\!\searrow\hspace{-10pt}\nwarrow\!\!|\!\!+\!\!|\!\!\swarrow\hspace{-10pt}\nearrow\!\swarrow\hspace{-10pt}\nearrow\!\rangle\langle\!\swarrow\hspace{-10pt}\nearrow\!\!\swarrow\hspace{-10pt}\nearrow\!\!|)\ {\rm for}\ \mu_y\!=\!\mu_z\!=\!0, \mu_x\!=\!\frac{1}{4} ,\ \ \ \ \ 
\label{ccs2}
\end{eqnarray}  
Here $|\!\updownarrow\rangle,|\!\leftrightarrow\rangle$ and $|\searrow\hspace{-10pt}\nwarrow\rangle,|\swarrow\hspace{-10pt}\nearrow\rangle$ denote the vertical, horizontal, diagonal and antidiagonal polarisation states respectively. The states (\ref{mes1}) and (\ref{mes2}) are entangled while the states (\ref{ccs1}) and (\ref{ccs2}) are classically correlated. The coincidence detection probability for (\ref{BellDiag}) is given  by
\begin{equation}
P_{BD}=\frac{1}{4}+\mu_x\sin(2\theta_1)\sin(2\theta_2)+\mu_z\cos(2\theta_1)\cos(2\theta_2).
\label{coincdet}
\end{equation}

\begin{figure}[!t]
\begin{center}
\includegraphics[scale=0.6]{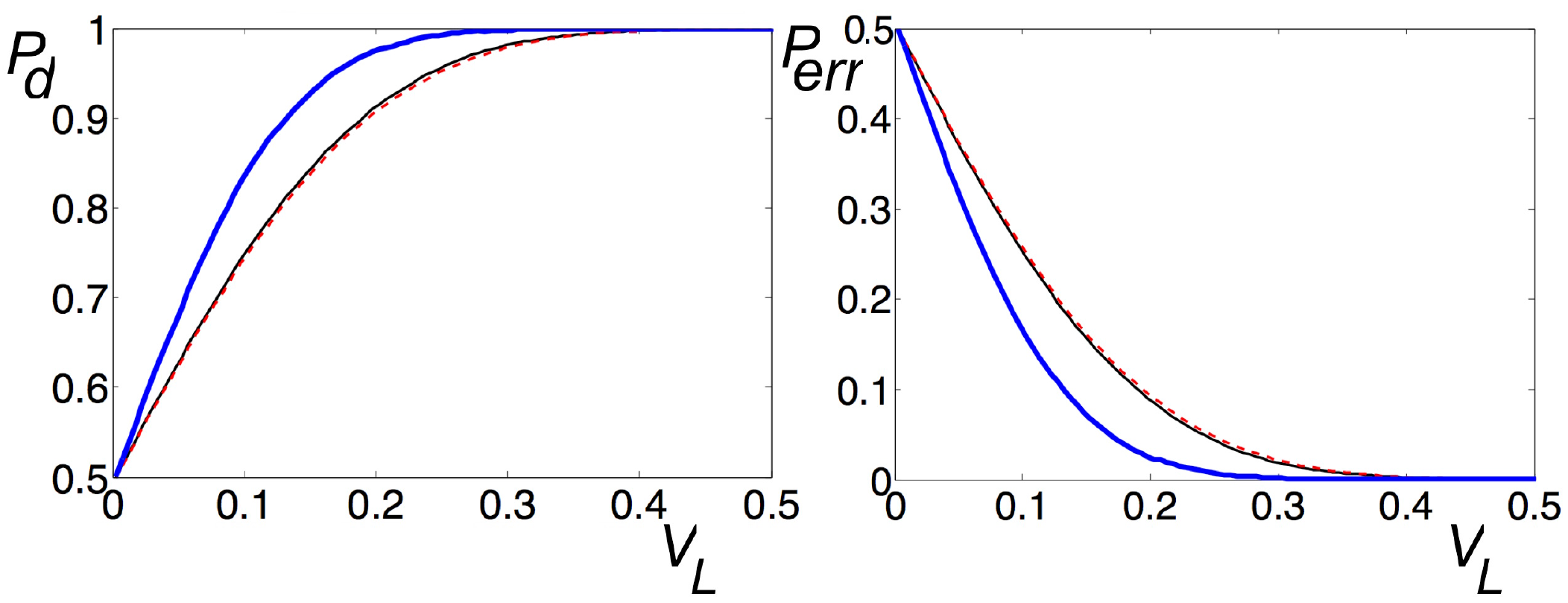}
\caption{(Color online.) Probability of intrusion detection (left panel) and false alarm probability (right panel) as a function of the jamming level 
 (\ref{dworst}). The noise variance is $\sigma=0.1$. The thin black, red dashed and thick blue lines correspond to the legitimate imagers using two classical states  (\ref{ccs1}) and (\ref{ccs2}), one entangled and one classical (\ref{mes1}) and (\ref{ccs1}) and two entangled states \ref{mes1}) and (\ref{mes2}) respectively. For $V_L>0.5$ there is no significant difference between imaging with classical and quantum states.}
\label{probdetworst}
\end{center}
\end{figure}

The visibility difference $d$ (\ref{dworst}) and the detection probability (\ref{pede}) depend on the intruder's set of possible states. We assume that this set is given by (\ref{BellDiag}). For pairs of states from the set (\ref{mes1})-(\ref{ccs2}) the probability of detection is plotted in the left panel of Fig.~\ref{probdetworst} in the worst case scenario for different values of the level of jamming (\ref{loj}), while the probability of false detection is plotted in the right panel. The value of the noise variance is chosen to be $\sigma=0.1$.
For pairs of entangled states 
the detection probability is larger than for pairs of classically correlated states. In particular, for a level of jamming $V_L=0.1$ the probability of jamming detection is about $0.82$ when two classical states are used for legitimate imaging while $P_d=0.92$ if entangled states are used. Preliminary numerical calculations show that the situation does not change in the intruder's favour if the intruder changes local bases of states (\ref{BellDiag}). However, this analysis requires further investigation. \vspace{3pt}
   
\begin{figure}[!t]
\begin{center}
\includegraphics[scale=0.2]{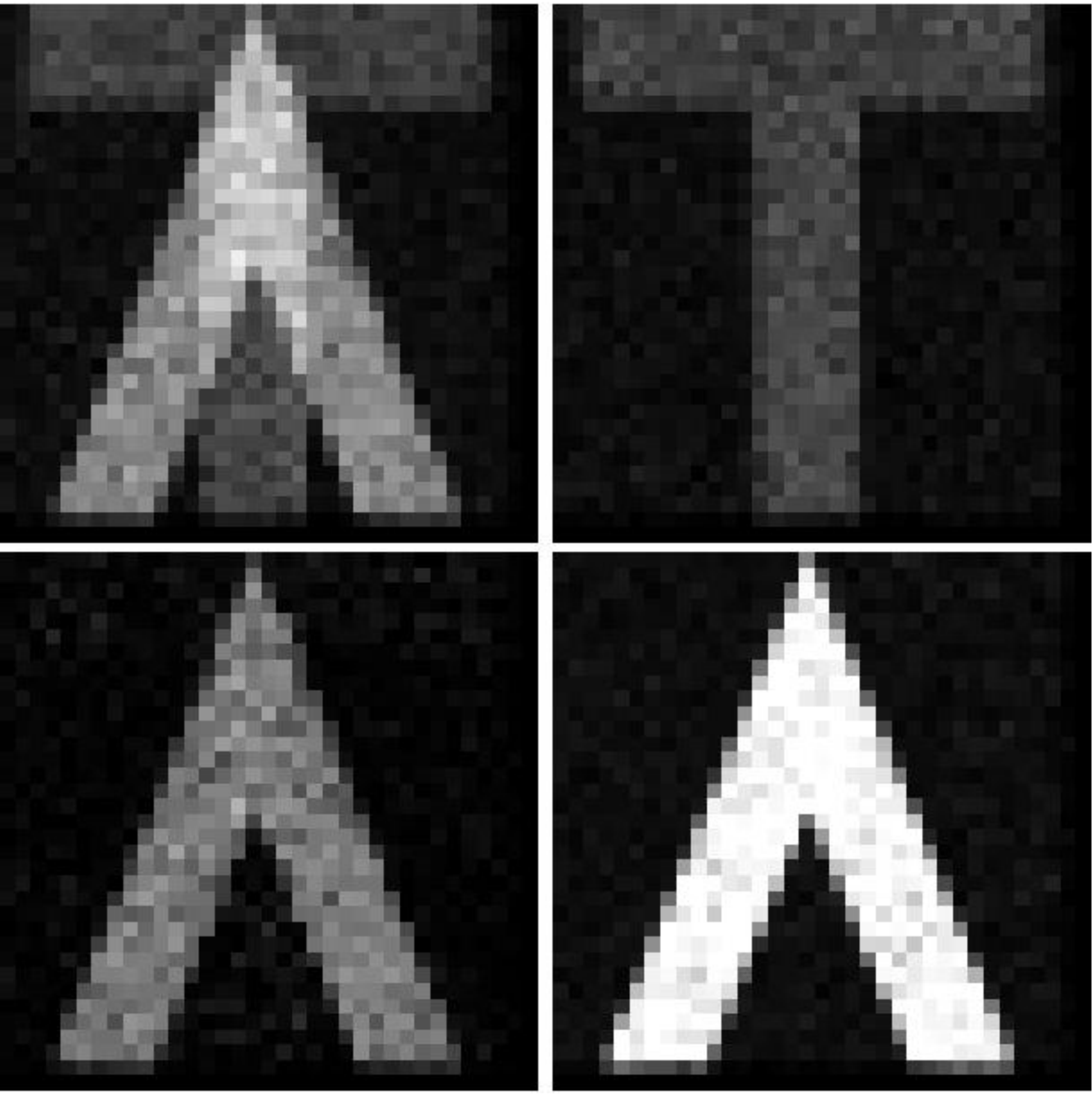}
\caption{The upper left image shows the mixture of the correct ($\Lambda$-shape) and the false ($T$-shape) image obtained using states (\ref{mes1}) and (\ref{ccs1}) respectively. The upper right image shows the false image. In this case, the correct part disappears because the legitimate states (\ref{mes2}) are blocked by the polarisers. The lower left image shows the recovered image. For comparison, the lower right image shows the image without jamming.}
\label{shadows}
\end{center}
\end{figure}


\subsection{Jamming detection probability and recovery protocol - example} A simulation of an example of the recovery protocol is illustrated in figure \ref{shadows}. Here, we assume that in order to image a correct $\Lambda$-shaped object legitimate imagers use two classes of photons characterised by maximally entangled  polarisation states $|\psi_1\rangle$ and $|\psi_2\rangle$ given by (\ref{mes1}) and (\ref{mes2}) respectively.
During this process intercept-resend jamming occurs with intercepting rate $r=0.5$. A false $T$-shaped image is imposed on the correct one. In this example we study a classical attack in which the false image is formed by photons in a classically correlated polarisation state $\rho^E=\omega_1$ given in (\ref{ccs1}).
The number of photons used in the simulation in order to form one image in a unit of time is $n=10^5$. We assume that the detectors that are used for imaging record accidental counts, dark counts and other events that contribute to general noise at rate $10^4$ for the unit of time across all pixels, i.e. $10\%$ of the number of photons. 

The legitimate imagers chose polariser angles $\theta_1=\theta_2=\pi/4$. The probability of the photons being detected are given by Eq. (\ref{coincdet}). This choice of angles provides the maximum state dependent visibility for imaging without jamming
\begin{equation}
V(\rho_1,\rho_2,\{\theta_i\})=1,
\end{equation}
where $\rho_1=|\psi_1\rangle\langle\psi_1|$ and $\rho_2=|\psi_2\rangle\langle\psi_2|$. The maximum value means that a correct image completely disappears when the state of the imaging photons is changed from $\rho_1$ to $\rho_2$, as visualised in the upper panels of figure \ref{shadows}. 

To estimate the probability of jamming detection we also calculate the state dependent visibility for jammed states $\rho_1'=(1-r)\rho_1+r\rho^E$ and $\rho_2'=(1-r)\rho_2+r\rho^E$. Here we calculate this quantity for the central region of the image in which a part of the false `T' is superimposed on the correct `$\Lambda$'. The visibility is 
\begin{equation}
V(\rho_1',\rho_2',\{\theta_i\})=0.5
\end{equation}
and it can be measured experimentally. For an assumed variance of Gaussian noise $\sigma=0.1$ the difference between the visibility without intrusion and the one with jamming in units of $\sigma$ is $d=5$ which guarantees that we detect the intrusion almost with certainty ($P_d\approx 1$, see (\ref{pede})).

According to the recovery protocol the correct image is obtained as the difference between two images from the upper panels of figure \ref{shadows}. The recovered image is shown in the lower right panel of this figure, while in the lower left panel we show the correct image obtained without jamming. 

During the imaging process random external noise independent of the image is recorded by detectors. We can observe that the noise level is reduced as an effect of our recovery protocol. Let us compare the levels of noise on two lower images of figure \ref{shadows}. The average amount of dark counts per pixel is about $8.7$ for the chosen unit of time in the lower right panel. The level of noise in the case of the recovered image (the lower left panel) is about $5$. The ratio between the average brightness of the $\Lambda$-shaped area and the noise without jamming is about $14$ while for the recovered image this ratio is about $12$. In consequence, the reduction of the absolute level of noise is observed in the simulation as an effect of the recovery protocol. The relative noise is increased due to a difference in brightness between the images observed with or without jamming.

\section{Concluding remarks and extensions}\label{sec6} 
Imaging in which an imager controls the signal used to test a remote object is more robust than imaging relying on the signal coming from the object or from an uncontrollable source. This control gives us the ability to detect jamming and eliminate the jammed part of the image. We describe a detection and recovery protocol relying on the control of the polarisation of light. The protocol can be applied to negate the effect of jamming by a malevolent party as well as reducing the impact of background noise and signals on the imaging. Security of this protocol is provided by the fundamental fact that the party who prepares quantum states has more information about the states than it is possible to extract from a measurement. This creates an informational advantage of the legitimate imager with respect to an intruder. The latter cannot perfectly correlate the false image carrying states of light with the states from a legitimate source. As a consequence, the informational advantage can be used for false image detection and correct image restoration, as is done in our protocol.

In the description of our protocol we assumed that the state of the intruder is stationary during the imaging process, at least on average. This is a reasonable consequence of the assumption that the intruder does not distinguish between states from the legitimate source and create correlations with the changes of these states. Let us consider the situation in which this assumption is relaxed allowing for partial correlations between the intruder's states and the states from the legitimate source. Even in this case we are able to recognise the false contribution. It is enough to observe an image that appears when a filter at the legitimate detector completely blocks one of the legitimate states. This situation is shown in the upper right panel of fig. \ref{shadows}. The false image contribution still appears there. In this case, however, we cannot use simple difference between two images corresponding to different legitimate states to recover the correct image, since the brightness of the false image is now different on each of them. On the other hand, a modification of the protocol, taking a weighted difference between the two images is still possible, which will allow us to correct for this problem. 

Free space application of the light beams together with the necessity of preservation of (or at least controlled interaction with) their polarisation states during imaging influences the conditions under which the protocol is applicable. In particular, the distance of the imaging is limited. Therefore, protocols like this one may not necessarily be considered in the context of radar type imaging of remote unknown objects. Instead, as the robots and drones industry is rapidly developed \cite{Joel2013} they can be parts of modern surveillance systems with application of such proxy observers sent to vulnerable places of an area being protected. In this example readout of the device is more trusted if controlled states of light are used instead of the signal sent from the device in the border region between protected and uncontrollable areas. The protocol can be also used when the robot is sent to view an object in a noisy environment, where direct imaging is impossible. Another range of possible applications appears in the context of the general free space optical communication~\cite{LopezMartinez2015}. 

Correlation imaging can provide further information. For example, due to timing information we can confirm if beam path lengths are equal or have been changed as a consequence of jamming. We have shown here a further example of the advantage to be gained by using quantum over classically correlated states in coincidence ghost imaging.  

We have discussed the situation in which jamming photons were directed to the detectors through the analysing polarisers. One alternative available to the intruder is jamming in which these photons are sent directly to the detectors side-stepping the polarisation analysers. Then the probability of detection does not depend on the polarisation states. As the intruder cannot optimize over these states the probability of intrusion detection is larger than in the case of jamming through the analysers, so this is not a useful intrusion strategy.\vspace{3pt}


\noindent
{\bf Acknowledgements}
The work was supported by the QuantIC Project of the UK Engineering and Physical Sciences Research Council (EP/M01326X/1). The authors thank Mehul Malik for useful discussion of an early version of this work and Masahide Sasaki for pointing out possible applications.


\end{document}